\documentclass[twocolumn,showpacs,preprintnumbers,amsmath,amssymb,aps]{revtex4}

\usepackage{graphicx}
\usepackage{color}
\usepackage{amsmath,amssymb,amssymb}

\begin{document}

\noindent Published in: J. Stat. Mech. (2015) P06029

\title{ Inverse participation ratio and localization in topological insulator phase transitions }

\author{M. Calixto}
\affiliation{Departamento de Matem\'atica Aplicada and
Instituto Carlos I de F{\'\i}sica Te\'orica y
Computacional, Universidad de Granada,
Fuentenueva s/n, 18071 Granada, Spain\\ calixto@ugr.es}

\author{E. Romera}
\affiliation{Departamento de F\'{\i}sica At\'omica, Molecular y Nuclear and
Instituto Carlos I de F{\'\i}sica Te\'orica y
Computacional, Universidad de Granada, Fuentenueva s/n, 18071 Granada,
Spain\\ eromera@ugr.es}

\begin{abstract}
Fluctuations of Hamiltonian eigenfunctions, measured by the inverse participation ratio (IPR), turn out to characterize topological-band insulator transitions occurring 
in 2D Dirac materials like silicene, which is isostructural with graphene but with a strong spin-orbit interaction. Using monotonic properties of the IPR, as a function of a 
perpendicular electric field (which provides a tunable band gap), we define topological-like quantum numbers that take different values in the topological-insulator 
and band-insulator phases. 
\end{abstract}

\maketitle

\section{Introduction}

In the last years, the concepts of insulator and metal have been revised and a new category called ``topological insulators'' has emerged. 
In these materials, the energy gap $\Delta$ between the occupied and empty states is  inverted or ``twisted'' for surface or edge states 
basically due to a strong spin-orbit interaction $\Delta_\mathrm{so}$ (namely, $\Delta_\mathrm{so}=4.2$ meV for silicene). Although ordinary band insulators can also support conducting metallic states on the surface, 
topological surface states have special features due to symmetry protection, which make them immune to scattering from ordinary defects and can carry electrical 
currents even in the presence of large energy barriers.

The low energy electronic properties of a large family of topological insulators and superconductors are well described by the Dirac equation  \cite{Shen}, in particular, 
some 2D gapped  Dirac materials isostructural with graphene like: silicene, germanene, stannene, etc. Compared to graphene, these materials display a large 
spin-orbit coupling and show quantum spin Hall effects \cite{Kane,Bernevig}. Applying a perpendicular electric field $\mathcal{E}_z=\Delta_z/l$ ($l$ is the inter-lattice distance of the buckled honeycomb structure, namely 
$l=0.22$\AA\  for silicene) to the material sheet, generates a tunable band gap (Dirac mass) $\Delta_{s}^{\xi}=(\Delta_z-s\xi\Delta_{\mathrm{so}})/2$ ($s=\pm 1$ and $\xi=\pm 1$ denote spin and valley, respectively).
There is  a topological phase transition  \cite{tahir2013}  from a topological  
insulator (TI, $|\Delta_z|<\Delta_\mathrm{so}$) to a band insulator (BI, $|\Delta_z|>\Delta_\mathrm{so}$), at a charge neutrality
point (CNP)  $\Delta_z^{(0)}=s\xi\Delta_\mathrm{so}$, where there is a gap cancellation  between the perpendicular electric field
and the spin-orbit coupling, thus exhibiting the aforementioned  semimetal behavior. In general, a TI-BI transition is characterized by a band inversion with
a level crossing at some critical value of a control parameter (electric field, quantum well thickness  \cite{Bernevig},
etc).

Topological phases are characterized by topological charges like Chern numbers. For an insulating state $|\psi({k})\rangle$, a ``gauge potential'' 
$a_j({k})=-i\langle\psi({k})|\partial_{k_j}\psi({k})\rangle$ can be defined in momentum space $(k_x,k_y)$, so that the Chern number $C$ 
is the integral $C=\int d^2k f({k})/2\pi$ of the Berry curvature 
$f=\partial_{k_x}a_y-\partial_{k_y}a_x$ over the first Brillouin zone. When the Hamiltonian is given by \eqref{hamiltoniano}, the Chern number is obtained 
as $C_s^\xi=-\xi\,\mathrm{sgn}(\Delta_s^\xi)/2$, so that a topological insulator phase transition (TIPT) occurs when the sign of the Dirac mass $\Delta_s^\xi$ changes \cite{Ezawarep}. 

In this article we propose the use of the inverse participation ratio (IPR) of Hamiltonian eigenvectors as a characterization of the topological phases TI and BI. The IPR measures the 
spread of a state $|\psi\rangle$ over a basis  $\{|i\rangle\}_{i=1}^N$. More precisely, if $p_i$ is the 
probability of finding the (normalized) state $|\psi\rangle$ in $|i\rangle$, then the IPR is defined as $I_\psi=\sum_i p_i^2$. If $|\psi\rangle$ only ``participates'' of a single state $|i_0\rangle$, then 
$p_{i_0}=1$ and $I_\psi=1$ (large IPR), whereas if $|\psi\rangle$ equally participates on all of them (equally distributed), $p_{i}=1/{N}, \forall i$, then $I_\psi=1/N$ (small IPR). 
Therefore, the IPR is a measure of the localization of $|\psi\rangle$ in the corresponding basis. Equivalently, the (R\'enyi) entropy $S_\psi=-\ln I_\psi$ is a measure of the delocalization of 
$|\psi\rangle$. We shall see that electron and hole IPR curves cross at the CNP, as a function of the electric field $\Delta_z$, and the crossing IPR value turns out to be an universal quantity, independent of Hamiltonian parameters. 
Moreover, the different monotonic character (slopes' sign) of combined (electron plus holes) IPRs in the TI and BI regions turn out to characterize both phases.

These and related information theoretic and statistical measures have proved to be useful in the description and characterization of quantum phase transitions (QPTs) of several 
paradigmatic models like: Dicke model of atom-field interactions \cite{romera11,JSM,romera12,calixto12,husidi}, vibron model of molecules \cite{husivi,JPAvibron,PRAcoupledbenders} and 
the ubiquitous Lipkin-Meshkov-Glick model \cite{LMG,EPL}. An important difference between QPT and TIPT is that the first case entails an abrupt symmetry change and the second one 
does not (necessarily). However, the use of information theoretic measures, like Wehrl entropy \cite{calixto14} and uncertainty relations \cite{romera14},  has proved to 
be useful to characterize TIPTs too. Moreover, we must also say that strong fluctuations of eigenfunctions (characterized by a set IPRs) also represent one of
the hallmarks of the traditional Anderson metal-insulator transition \cite{Anderson,Wegner,Brandes,Evers,Aulbach}. In fact, 
the phenomenon of localization of the electronic wave function can be regarded as the key manifestation of quantum coherence at a macroscopic scale in a 
condensed matter system.

The paper is organized as
follows. Firstly, in Section \ref{Hamilsec}, we shall introduce the low energy Hamiltonian describing the electronic properties of 
some 2D Dirac materials like silicene, germanene, stantene, etc, in the presence of perpendicular electric and magnetic  fields. 
Then, in Section \ref{IPRsec}, we will compute the IPR  of Hamiltonian eigenvectors  and we shall discuss the (different) structure of IPR curves as a 
function of the electric field across TI and BI regions. Section \ref{conclusec} is devoted to final comments and conclusions.

\section{Low energy Hamiltonian}\label{Hamilsec}

The low energy dynamics of a large family of topological insulators (namely, honeycomb structures) is described by the Dirac Hamiltonian 
in the vicinity of the Dirac points $\xi=\pm 1$ \cite{Tabert2013} 
\begin{equation}
H_s^{\xi}=v (\xi\sigma_x  p_x+ \sigma_y  p_y )-\frac12\xi
s \Delta_{\mathrm{so}} \sigma_z+ \frac12\Delta_z \sigma_z,
\label{hamiltoniano}
\end{equation}
where ${\sigma}_j$ are the usual Pauli matrices, $v$ is the Fermi
velocity of the corresponding material (namely, $v=4.2\times 10^5$m/s for silicene) and $\Delta_{\mathrm{so}}$ and $\Delta_z$ are the spin-orbit and electric field couplings. 
The application of a perpendicular magnetic field $B$ is implemented through the  minimal coupling $\vec{p}\to \vec{p}+{e}\vec{A}$ for the momentum, 
where $\vec{A}=(-By,0)$ is the vector potential in the Landau gauge.  The Hamiltonian eigenvalues 
and eigenvectors at the $\xi$ points  are given by
 \cite{Tabert2013}
\begin{equation}
E_{n}^{s\xi}=\left\{\begin{array}{l} \mathrm{sgn}(n) \sqrt{|n|\hbar^2\omega^2 + (\Delta_{s}^{\xi})^2}, \quad n\neq 0, \\ 
              -\xi\Delta_{s}^{\xi}, \quad n= 0, \end{array}\right.\label{especteq}
            \end{equation}
 and
\begin{equation}
|n\rangle_{s}^{\xi}=\left(\begin{array}{c}
-i A_{n}^{s\xi}||n|-\xi_+\rangle\\
B_{n}^{s\xi}||n|-\xi_-\rangle
\end{array}\right),
\label{vectors}
\end{equation}
where we denote by $\xi_\pm=(1\pm\xi)/2$, the Landau level index $n=0,\pm 1,\pm 2,\dots$, the cyclotron frequency 
$\omega=v\sqrt{2eB/\hbar}$, the lowest band gap $\Delta_{s}^{\xi}\equiv(\Delta_z-s\xi\Delta_{\mathrm{so}})/2$ and  the coefficients 
$A_{n}^{s\xi}$ and $B_{n}^{s\xi}$ are given by \cite{Tabert2013}
\begin{eqnarray}
 A_{n}^{s\xi}&=&\left\{\begin{array}{l}
 \mathrm{sgn}(n)\sqrt{\frac{|E_{n}^{s\xi}|+\mathrm{sgn}(n)\Delta_{s}^{\xi}}{{2|E_{n}^{s\xi}|}}},  \quad n\neq 0, \\
\xi_-,  \quad n=0, 
\end{array}\right.\nonumber\\ 
 B_{n}^{s\xi}&=&\left\{\begin{array}{l}
\sqrt{\frac{|E_{n}^{s\xi}|-\mathrm{sgn}(n)\Delta_{s}^{\xi}}{{2|E_{n}^{s\xi}|}}},  \quad n\neq 0, \\
\xi_+,  \quad n=0, 
\end{array}\right.\label{coef}
\end{eqnarray}
The vector  $||n|\rangle$ denotes an orthonormal Fock state of the harmonic
oscillator. We are discarding a trivial plane-wave dependence $e^{ikx}$ in the $x$ direction that does not affect out IPR calculations, which only depend on the $y$ direction 
in this gauge.

\begin{figure}
\begin{center}
\includegraphics[width=8cm]{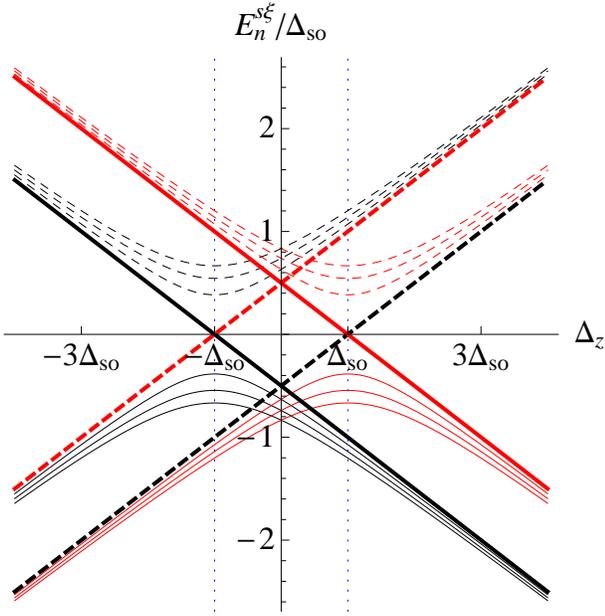}
\end{center}
\caption{Low energy spectra of silicene as a function of the external electric potential $\Delta_z$ for $B=0.01$ T. 
Landau levels $n=\pm 1, \pm 2$ and $\pm 3$, at valley $\xi=1$,  are represented by 
dashed (electrons) and solid (holes) thin lines, black for $s=-1$ and red for
$s=1$ (for the other valley we simply have $E_{n}^{s,-\xi}=E_{n}^{-s,\xi}$). The lowest Landau level $n=0$ is represented by thick lines at both valleys:  solid at $\xi=1$ and dashed 
at $\xi=-1$. Vertical blue dotted grid lines indicate the CNPs separating BI ($|\Delta_z|>\Delta_\mathrm{so}$) from TI   ( $|\Delta_z|<\Delta_\mathrm{so}$) phases.}
\label{energias}
\end{figure}

As already stated, there is a prediction (see e.g. \cite{drummond2012,liu2011,liub2011,Ezawa}) that when the gap 
$|\Delta_{s}^{\xi}|$ vanishes at the CNP $|\Delta_z|=\Delta_\mathrm{so}$, 
silicene undergoes a phase transition from a topological insulator (TI,  $|\Delta_z|<\Delta_\mathrm{so}$) to a band insulator (BI, $|\Delta_z|>\Delta_\mathrm{so}$). 
This topological phase transition entails an energy band inversion. Indeed, in Figure \ref{energias} we show the low energy spectra \eqref{especteq} 
as a function of the external electric potential $\Delta_z$ for $B=0.01$ T. One can see that there is a band inversion 
for the $n=0$ Landau level (either for spin up and down) at both valleys. The
energies $E_0^{1,\xi}$ and $E_0^{-1,\xi}$ have the same sign in the BI phase and different sign in the TI phase, thus distinguishing both regimes. 
We will provide an alternative description of this phenomenon in terms of IPR relations for the Hamiltonian eigenstates \eqref{vectors}, thus providing a quantum-information characterization of TIPT.

\section{IPR  and topological insulator phase transition} \label{IPRsec}

The IPR of a given state $I_\psi$ is related to a certain basis. In this article we shall chose the position representation to  
write the Hamiltonian eigenstates \eqref{vectors}. 
We know that Fock (number) states $|n\rangle$ can be written in position representation as
\begin{equation}
 \langle y|n\rangle= \frac{\omega^{1/4}}{\sqrt{2^n n!\sqrt{\pi}}}e^{-\omega y^2/2} H_n\left(\sqrt{\omega} y\right)
\end{equation}
where $H_n$ are the Hermite polynomials of degree $n$. The number-state density in position 
 space is $\rho_n(y)=|\langle y|n\rangle|^2$, which is normalized according to $\int \rho_n(y)dy=1$. 
Now, taking into account  Eq. (\ref{vectors}), the density for the Hamiltonian eigenvectors \eqref{vectors} in position representation is given by
\begin{equation}
 \rho_{n}^{s\xi}(y)=(A_{n}^{s\xi})^2|\langle y||n|-\xi_+\rangle_{s}^{\xi}|^2 +(B_{n}^{s\xi})^2|\langle y||n|-\xi_-\rangle_{s}^{\xi}|^2 .
\end{equation}
The IPR of a Hamiltoninan eigenstate in position representation is then calculated as
\begin{equation}
 I_{n}^{s\xi}\equiv \int_{-\infty}^{\infty} \left(\rho_{n}^{s\xi}(y)\right)^2 dy.\label{mommu}
\end{equation}
As a previous step, we need the following integrals of Hermite density products:
\begin{eqnarray}
 M_{n,m}&\equiv&\int_{-\infty}^{\infty}\rho_{n}(y)\rho_{m}(y)dy=\nonumber\\ &=&\sqrt{\frac{\omega}{2\pi}}\begin{pmatrix}
      1& \frac{1}{2} & \frac{3}{8} &\frac{5}{16} & \dots \\
      \frac{1}{2} &\frac{3}{4} & \frac{7}{16} &  \frac{11}{32}  & \dots \\
      \frac{3}{8} &\frac{7}{16} & \frac{41}{64} & \frac{51}{128}  & \dots \\
      \frac{5}{16} & \frac{11}{32} & \frac{51}{128} &\frac{147}{256} & \dots \\
      \vdots &  \vdots & \vdots & \vdots & \ddots
     \end{pmatrix}, 
\end{eqnarray}
for $n,m=0,1,2,3\dots$ We shall also restrict ourselves to the valley $\xi=1$, omitting this index  from 
\eqref{vectors} and \eqref{coef}. All the results for the valley $\xi=1$ are straightforwardly translated to the valley $\xi=-1$ by swapping 
electrons for holes (i.e., $n\leftrightarrow -n$) and spin up for down (i.e., $s\leftrightarrow -s$). Taking into account all these considerations, the IPR of a Hamiltonian eigenstate 
is explicitly written as
\begin{eqnarray}
 I_{n}^{s}&\equiv& 
 (A_{n}^{s})^4 M_{|n|-1,|n|-1}+(B_{n}^{s})^4
M_{|n|,|n|}\nonumber\\ && +2(A_{n}^{s}B_{n}^{s})^2 M_{|n|,|n|-1}.
\end{eqnarray}

\begin{figure}
\includegraphics[width=8cm]{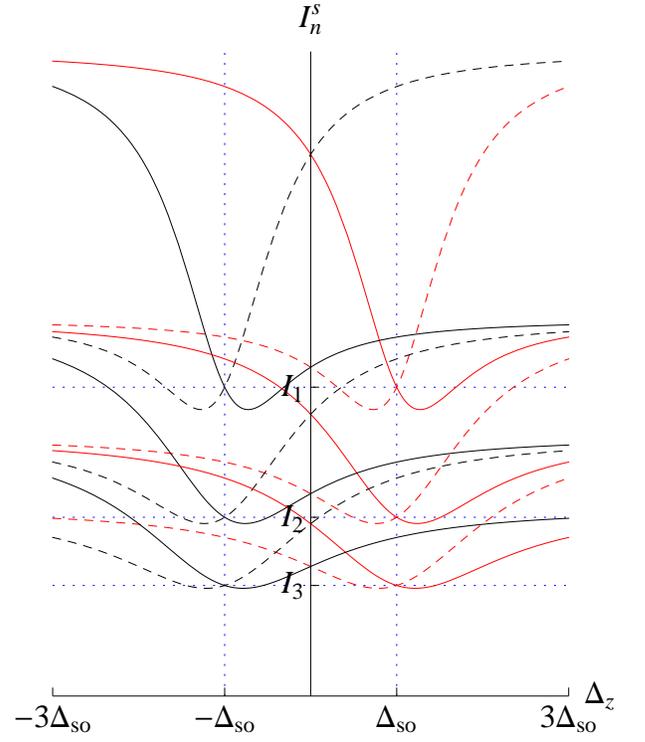}
\caption{ \label{IPRpos} IPR $I_{n}^{s\xi}$ in position space at  valley $\xi=1$
as a function of the electric potential $\Delta_z$ for the Landau levels: $n=\pm 1, \pm 2$ and $\pm 3$ and magnetic field $B=0.01$T. 
Electron IPR curves are dashed and hole curves are solid, black for spin down  $s=-1$ and red for spin up $s=1$. Electron and hole IPR curves 
cross  at the critical value of the electric potential $\Delta_z^{(0)}=s\Delta_{\mathrm{so}}$; vertical blue dotted grid lines
indicate this CNPs and horizontal blue dotted grid lines indicate the crossing IPR values \eqref{crossingIPR}.}
\end{figure}

\begin{figure}
\begin{center}
\includegraphics[width=8cm]{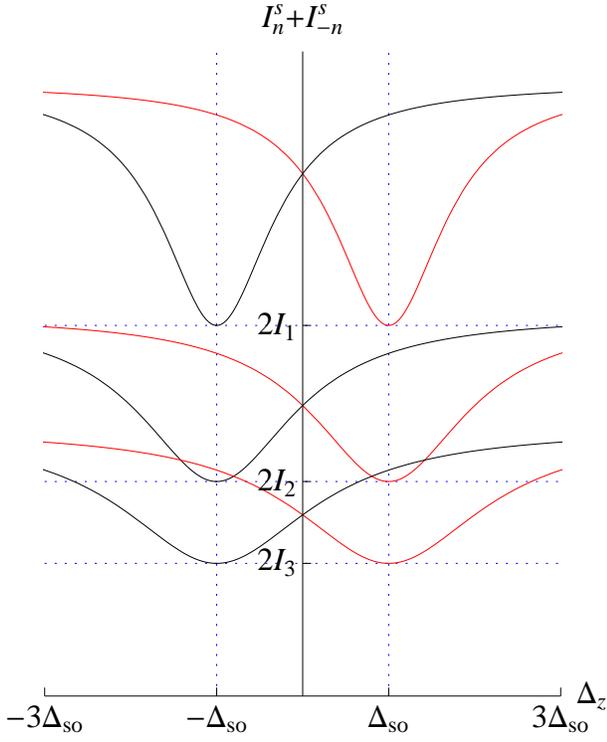}
\end{center}
\caption{ \label{IPRsumpos} Combined IPR $I_{n}^{s\xi}+I_{-n}^{s\xi}$ of electron plus holes  at  valley $\xi=1$
as a function of the electric potential $\Delta_z$ for the Landau levels: $n=\pm 1, \pm 2$ and $\pm 3$ and magnetic field $B=0.01$T. Black 
curves for   $s=-1$ and red for  $s=1$. Vertical blue dotted grid lines
indicate the CNPs and horizontal blue dotted grid lines indicate the minimum combined IPR values.}
\end{figure}

\begin{figure}
\begin{center}
\includegraphics[width=8cm]{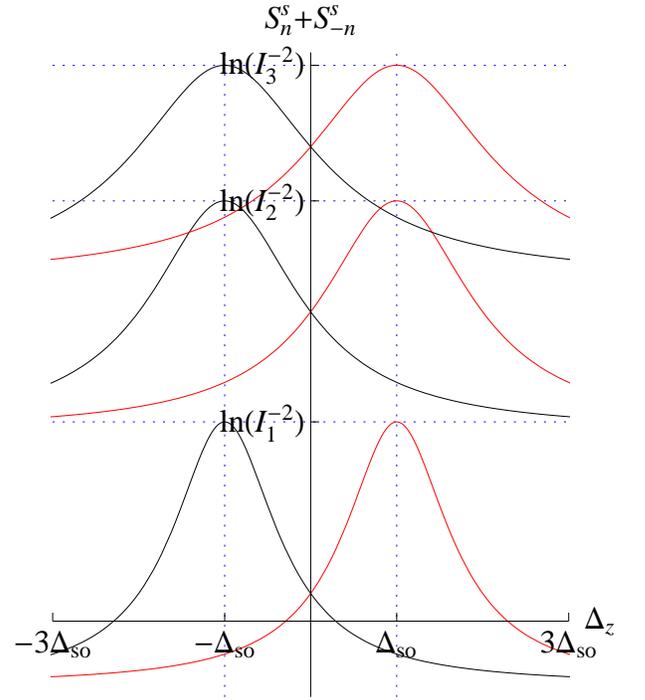}
\end{center}
\caption{ \label{Entrosumpos} Combined entropies $S_{n}^{s\xi}+S_{-n}^{s\xi}$ ($S=-\ln I)$ of electron plus holes  at  valley $\xi=1$
as a function of the electric potential $\Delta_z$ for the Landau levels: $n=\pm 1, \pm 2$ and $\pm 3$ and magnetic field $B=0.01$T. Same color code as in Fig. \ref{IPRsumpos}. 
Minimum entropies are attained at the CNP and are indicated by horizontal blue gris lines.}
\end{figure}

In Figure  \ref{IPRpos} we plot $I_{n}^s$ as a function of the external electric potential
$\Delta_z$ for the Landau levels: $n=\pm 1, \pm 2$ and $\pm 3$. 
 Electron and hole IPR curves cross at the CNPs $\Delta_z^{(0)}=s\Delta_\mathrm{so}$, where they take the values:
 \begin{equation}
  I_1=\sqrt{\frac{\omega}{2\pi}}\frac{11}{16},\,  I_2=\sqrt{\frac{\omega}{2\pi}}\frac{145}{256},\,  I_3=\sqrt{\frac{\omega}{2\pi}}\frac{515}{1024}.\label{crossingIPR}
 \end{equation}
Note that, except for $\omega$, the critical crossing IPR value $I_n$ only depends on the Landau level $n$, and not on any other physical magnitude, thus 
providing a universal characterization of the topological insulator transition.
We have checked that the smaller the magnetic field strength, the greater the slope of the electron-hole IPR curves at the CNP. The asymptotic values 
$I_n({\pm\infty})=\lim_{\Delta_z\to\pm\infty}I_n^s(\Delta_z)$ also exclusively depend on  $n=1,2,\dots$ and $\omega$, and are:
 \begin{equation}
  I_1(\infty)=\sqrt{\frac{\omega}{2\pi}},\,  I_2(\infty)=\sqrt{\frac{\omega}{2\pi}}\frac{3}{4},\,  I_3(\infty)=\sqrt{\frac{\omega}{2\pi}}\frac{41}{64},\dots\label{asymptoticIPR}
 \end{equation}
fulfilling $I_{-n}(-\infty)=I_{n}(\infty)$ and $I_{-|n|}(\infty)=I_{|n|+1}(\infty)$.

The crossing of IPR curves for electron and holes characterizes the CNPs. However, in order to properly characterize TI and BI phases, the combined IPR of electrons plus holes,  
$Y_n^s=I_{n}^{s}+I_{-n}^{s}$, offers a better indicator of the corresponding transition (see Figure \ref{IPRsumpos}). 
Indeed, on the one hand, $Y_n^s$ display global minima at the CNPs (highly 
delocalized states); on the other hand, the quantity  
\begin{equation}
C(\Delta_z)=\mathrm{sgn}\left(\frac{\partial Y_s}{\partial \Delta_z} \frac{\partial Y_{-s}}{\partial \Delta_z}\right),\label{charge1}
\end{equation}
plays the role of a ``topological charge'' (like a Chern number) so that 
\begin{equation}
 C(\Delta_z)=\left\{\begin{array}{ccl} 1,& |\Delta_z|>\Delta_\mathrm{so}& \mathrm{BI}, \\ -1,& |\Delta_z|<\Delta_\mathrm{so}& \mathrm{TI}. \end{array}\right.
\end{equation}
That is, the sign of the product of combined IPR slopes for spin up and down, clearly characterizes the two (TI and BI) phases. 

\begin{figure}
\includegraphics[width=8cm]{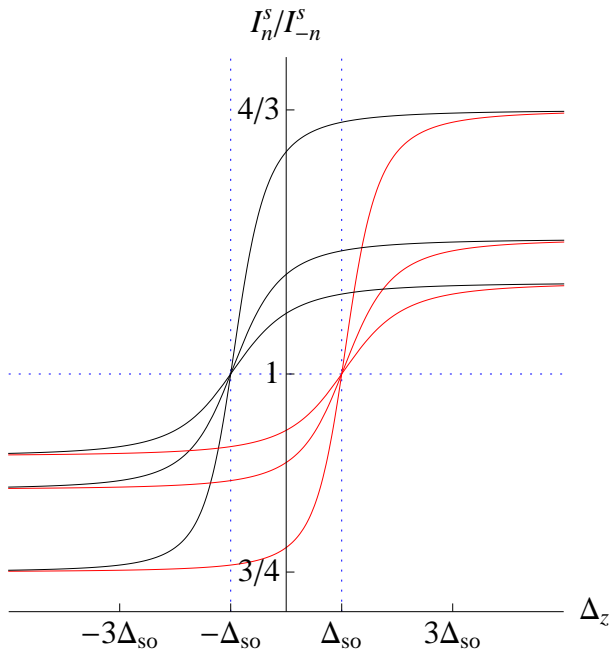}
\caption{ \label{IPRquotpos} Electron-hole IPR quotients  $I_{n}^{s\xi}/I_{-n}^{s\xi}$ at  valley $\xi=1$
as a function of the electric potential $\Delta_z$ for the Landau levels: $n=\pm 1, \pm 2$ and $\pm 3$ and magnetic field $B=0.01$T. Same color code as in Fig. \ref{IPRsumpos}. 
Electron-hole IPR ratios are increasing functions of $\Delta_z$.}
\end{figure}

The product $I_{n}^{s}\times I_{-n}^{s}$ of electron times hole IPRs also exhibits a critical behavior at the CNPs and characterizes both phases. Actually, the combined entropy ($S=-\ln I$) 
\begin{equation}
 S_{n}^{s}+S_{-n}^{s}=-\ln(I_{n}^{s}\times I_{-n}^{s}), 
\end{equation}
is minimum at the CNPs (highly delocalized states), as can be appreciated in Figure \ref{Entrosumpos}. For the quotient $Q_n^s\equiv I_{n}^{s}/I_{-n}^{s}$ we have that the 
quantity
\begin{equation}
\mathrm{sgn}\left(\frac{Q_n^s(\Delta_z)-1}{Q_n^{-s}(\Delta_z)-1}\right)=\left\{\begin{array}{ccl} 1,& |\Delta_z|>\Delta_\mathrm{so}& \mathrm{BI}, 
\\ -1,& |\Delta_z|<\Delta_\mathrm{so}& \mathrm{TI}, \end{array}\right.\label{charge2}
\end{equation}
also characterizes both phases, as can be explicitly appreciated in Figure \ref{IPRquotpos}.

\section{Conclusions}\label{conclusec}

We have studied localization properties of the Hamiltonian eigenvectors $|n\rangle_s^\xi$ ($n,\xi,s$ denote: Landau level, valley and spin, respectively) for 
2D Dirac materials isostructural with graphene (namely, silicene) in the presence of perpendicular magnetic $B$ and electric $\Delta_z$ fields. The electric field 
provides a tunable band gap $\Delta_s^\xi=(\Delta_z-s\xi\Delta_\mathrm{so})/2$ which is ``twisted'' at the charge neutrality points (CNPs)  $|\Delta_z|=\Delta_\mathrm{so}$, 
for surface states, due to a strong spin-orbit interaction $\Delta_\mathrm{so}$. The topological insulator (TI) and band insulator (BI) phases are then characterized by 
$|\Delta_z|<\Delta_\mathrm{so}$ and $|\Delta_z|>\Delta_\mathrm{so}$, respectively, or by the sign of $\Delta_s^\xi$ (the Chern number). 

We have proposed information-theoretic measures, based on the inverse participation ratio $I_n^{s\xi}$ (IPR), as alternative signatures of a topological insulator phase transition. 
The IPR measures the localization of a state in the corresponding basis (position representation in our case). 
We have seen that IPR curves $I_n^{s\xi}(\Delta_z)$ of electrons ($n>0$) and holes ($n<0$) cross at the CNPs, the crossing value being a universal quantity  basically depending 
on the Landau level $n$. The combined IPR $Y_n^{s\xi}=I_n^{s\xi}+I_{-n}^{s\xi}$ of electrons plus holes is minimum at the CNPs; 
i.e. the combined state is highly delocalized (maximum entropy) at the transition point. The different monotonic behavior of combined $Y_n^{s\xi}$ and quotient 
$Q_n^s\equiv I_{n}^{s}/I_{-n}^{s}$ IPR curves across BI and TI regions provides topological (Chern-like) numbers \eqref{charge1} and \eqref{charge2} which characterize 
both phases.

Therefore, as it already happens for the traditional Anderson metal-insulator transition, we have shown that fluctuations of Hamiltonian eigenfunctions (characterized by IPRs) also 
describe topological-band insulator transitions.

\section*{Acknowledgments}
  The work was supported by 
the  Spanish projects:  MINECO FIS2014-59386-P,  CEIBIOTIC-UGR PV8 and the Junta de Andaluc\'{\i}a projects FQM.1861 and FQM-381.

\end{document}